\documentclass[twocolumn,showpacs,preprintnumbers,amsmath,amssymb]{revtex4}

\usepackage{graphicx,dcolumn,bm,times}

\begin{document}

\title{Excitable Greenberg-Hastings cellular automaton model on scale-free networks}

\author{An-Cai Wu,$^{1}$ Xin-Jian Xu,$^{2}$ and Ying-Hai Wang$^{1}$}

\affiliation{$^{1}$Institute of Theoretical Physics, Lanzhou University, Lanzhou Gansu 730000, China\\
$^{2}$Departamento de F\'{i}sica da Universidade de Aveiro,
3810-193 Aveiro, Portugal}

\date{\today}

\begin{abstract}
We study the excitable Greenberg-Hastings cellular automaton model
on scale-free networks. We obtained analytical expressions for no
external stimulus and the uncoupled case. It is found that the
curves, the average activity $F$ versus the external stimulus rate
$r$, can be fitted by a Hill function, but not exactly, and there
exists a relation $F\propto r^\alpha$ for the low-stimulus
response, where Stevens-Hill exponent $\alpha$ ranges from $\alpha
= 1$ in the subcritical regime to $\alpha = 0.5$ at criticality.
At the critical point, the range reaches the maximal. We also
calculate the average activity $F^{k}(r)$ and the dynamic range
$\Delta^{k}(p)$ for nodes with given connectivity $k$. It is
interesting that nodes with larger connectivity have larger
optimal range, which could be applied in biological experiments to
reveal the network topology.
\end{abstract}


\maketitle

Many systems in the real world, either naturally evolved or
artificially designed, are organized in a network fashion \cite
{Albert}. The most studied networks are the exponential networks
in which the nodes' connectivity distribution (the probability
$P(k)$ that a node if connected to other \emph{k} nodes) is
exponentially bounded. A typical model of this network is the
Watts-Strogatz (WS) graph \cite {Watts} which exhibits the \lq\lq
small-world\rq\rq phenomenon that was observed in realistic
networks \cite {Watts_1}. On the contrary, it has been observed
recently that the nodes' degree distribution of many networks have
power-law tails (scale-free property, due to the absence of a
characteristic value for the degrees) $P(k)\sim k^{- \gamma}$ with
$2 \leq \gamma \leq 3$ \cite {Barabasi}.

Scale-free (SF) networks have been widely studied in the past few
years, mainly because of their connection to a lot of real-world
structures, including networks in nature, such as the cell,
metabolic networks and the food web, artificial networks such as
the Internet and the WWW, or even social networks, such as sexual
partnership networks \cite {Albert}. The SF network is
characterized with the power-law degree distribution. Thus there
always exists a small number of nodes which are connected to a
large number of other nodes in SF networks. This heterogeneity
leads to intriguing properties of SF networks. A lot of work has
been devoted in the literature to the study of static properties
of the networks, while many interests are growing for dynamical
properties on these kind of networks. In the context of
percolation, SF networks are stable against random removal of
nodes while they are fragile under intentional attacks targeting
on nodes with high degree \cite {Albert2, Cohen}. Also, there are
absences of epidemic thresholds in SF networks \cite {Pastor} and
of the kinetic effects in reaction-diffusion processes taking
place on SF networks \cite {GA04}.

In this paper, we will study the excitable Greenberg-Hastings
cellular automaton (GHCA) model \cite {Greenberg78} on the SF
network, especially we focus here on the Barab\'{a}si and Albert
(BA) graph \cite {Barabasi}. Due to experimental data which
suggest that some classes of spiking neurons in the first layers
of sensory systems are electrically coupled via gap junctions or
ephaptic interactions, the GHCA model is employed to model the
response of the sensory network to external stimuli in some recent
works. Two-dimensional deterministic cellular automaton model was
studied by computer simulations \cite{Copelli05}. Analytical
results have recently been obtained for the one-dimensional
cellular automaton model under the two-site mean-field
approximation \cite{Furtado06}. In ref. \cite{Kinouchi}, Kinouchi
and Copelli studied the GHCA model on Erd\"os-R\'enyi random graph
with stochastic activity propagation, and they found a new and
important result that dynamic range maximized at the critical
point, which is perhaps the first clear example of signal
processing optimization at a phase transition.

In the $n$-state GHCA model \cite{Greenberg78} for excitable
systems, the instantaneous membrane potential of the $i$-th cell
($i=1,\ldots,N$) at discrete time $t$ is represented by
$x_i(t)\in\{0,1,\ldots,n-1\}$, $n\geq 3$. The state $x_i(t)=0$
denotes a neuron at its resting (polarized) potential, $x_i(t)=1$
represents a spiking (depolarizing) neuron and
$x_i(t)=2,\ldots,n-1$ account for the afterspike refractory period
(hyperpolarization). There are two ways for the $i$-th element to
go from state $x_i(t) = 0$ to $x_i(t+1) = 1$: a) due to an
external signal, modelled here by a Poisson process with rate $r$
(which implies a transition with probability
$\lambda=1-\exp(-r\Delta t)$ per time step); b) with probability
$p$, due to a neighbor $j$ being in the excited state in the
previous time step. If $x_i(t)\geq 1$, then
$x_i(t+1)=(x_i(t)+1)\text{mod }n$, regardless of the stimulus. In
other words, the rules state that a neuron only spikes if
stimulated, after which it undergoes an absolute refractory period
before returning to rest. Time is discrete. We assume $\Delta t =
1$ ms which corresponds to the approximate duration of a spike and
is the time scale adopted for the time step of the model. The
number of states $n$ therefore controls the duration of the
refractory period (which corresponds to $n-2$, in ms). In the
biological context, $r$ could be related for example with the
concentration of a given odorant presented to an olfactory
epithelium \cite{Rospars00}, or the light intensity stimulating a
retina \cite{Deans02}. We shall refer to $r$ as the stimulus rate
or intensity.

BA graph is a kind of SF networks and can be constructed according
to ref. \cite {Barabasi}. Starting from a small number $m_{0}$ of
nodes, every time step a new vertex is added, with $m$ links that
are connected to an old node $i$ with probability
$\Pi_{i}=k_{i}/\sum_{j}k_{j}$, where $k_{i}$ is the connectivity
of the $i$th node. After iterating this scheme a sufficient number
of times, we obtain a network composed by $N$ nodes with
connectivity distribution $P(k) \sim k^{-3}$ and average
connectivity $\langle k \rangle =2m$. For this kind network, the
absence of a characteristic scale for the connectivity makes
highly connected nodes statistically significant, and induces
strong fluctuations in the connectivity distribution which cannot
be neglected.

Let $\rho_{t}(s)$ be the densities of neurons which are in state
$s$ at time $t$. We have the normalization condition,
$\sum_{s=0}^{n-1}\rho_{t}(s)=1$. Since the dynamics of the
refractory state is deterministic, the equations for $s\geq 2$ are
simply
\begin{eqnarray}
\label{eq:refrac}
\rho_{t+1}(2) & = & \rho_{t}(1) \nonumber \\
\rho_{t+1}(3) & = & \rho_{t}(2) \nonumber \\
\ & \vdots & \ \nonumber \\
\rho_{t+1}(n-1) & = & \rho_{t}(n-2) \; .
\end{eqnarray}
By imposing the stationarity condition, we have
\begin{equation}\label{recursion2}
\rho_{t}(0)=1-(n-1)\rho_{t}(1),
\end{equation}
Following ideas developed by Pastor-Satorras et al. \cite
{Pastor}, to take into account strong fluctuations in the
connectivity distribution, we consider the relative density
$\rho_{t}{^{k}}(1)$ of active nodes with given connectivity $k$ by
the following equation
\begin{equation}
\partial_t \rho_{t}{^{k}}(1) = -\rho_{t}{^{k}}(1) +\lambda \rho_{t}{^{k}}(0)
+kp(1-\lambda)\rho_{t}{^{k}}(0) \Theta(\rho_{t}(1)), \label{bamf}
\end{equation}
where $\Theta(\rho_{t}(1))$ is the probability that any given link
point to an active node and is assumed as a function of the total
density of exciting nodes $\rho_{t}(1)$. In the steady state,
$\rho_{t}(1)$ is just a function of $\lambda$ and $p$. Thus, the
probability $\Theta$ becomes an implicit function of $\lambda$ and
$p$. By imposing the stationarity condition, $\partial_t
\rho_{t}{^{k}}(1)=0$, we obtain
\begin{equation}
\rho {^{k}}(1)= \frac{\lambda+(1- \lambda)pk \Theta (\lambda,p)}{1
+(n-1)[\lambda+(1- \lambda)pk \Theta (\lambda,p)]}. \label{barhok}
\end{equation}
This set of equations show that the higher the node connectivity,
the higher the probability to be in a spiking state. This
inhomogeneity must be taken into account in the computation of
$\Theta(\lambda,p)$. Indeed, the probability that a link points to
a node with $q$ links is proportional to $qP(q)$. In other words,
a randomly chosen link is more likely to be connected to an
exciting node with high connectivity, yielding the relation
\begin{equation}
\Theta(\lambda,p)=\sum_k \frac{kP(k)\rho {^{k}}(1)}{\sum_q qP(q)}.
\label{theta1}
\end{equation}
Since $\rho {^{k}}(1)$ is on its turn a function of
$\Theta(\lambda,p)$, we obtain a self-consistency equation that
allows to find $\Theta(\lambda,p)$ and an explicit form for Eq.
(\ref{barhok}). Finally, we can evaluate the order parameter
(persistence) $\rho (1)$ using the relation
\begin{equation}
\rho(1)=\sum_kP(k)\rho {^{k}}(1). \label{barho}
\end{equation}
For the BA graph, the full connectivity distribution is given by
$P(k)= 2m^{2}k^{-3}$, where $m$ is the minimum number of
connection at each node. By noticing that the average connectivity
is $\langle k \rangle = \int_{m}^{\infty} k P(k) dk =2m$, Eq.
(\ref{theta1}) gives
\begin{equation}
\Theta(\lambda,p) = m\int_{m}^{\infty}
\frac{1}{k^2}\frac{\lambda+(1-\lambda)pk\Theta(\lambda,p)}{1+(n-1)[\lambda+(1-\lambda)pk\Theta(\lambda,p)]},
\label{theta2}
\end{equation}
which yields the solution
\begin{equation}
\Theta(\lambda,p) =
\frac{\lambda}{b}+(\frac{\lambda}{b}-\frac{1}{n-1})\frac{ma\Theta(\lambda,p)}{b}
\ln\frac{ma\Theta(\lambda,p)}{ma\Theta(\lambda,p)+b},
\label{theta20}
\end{equation}
where $a=(n-1)(1-\lambda)p$ and $b=1+(n-1)\lambda$. We can solve
the above equation to obtain the solution $\Theta(\lambda,p)$.
Theoretically, combing Eqs. (\ref{barhok}) , (\ref{barho}) and
(\ref{theta20}), we can obtain the final result of $\rho(1)$. In
some special cases the analytical expressions can be obtained.

i) no external stimulus, i.e., $\lambda=0$ or $r=0$. We can obtain
\begin{equation}
\Theta(0,p) = \frac{e^{-1/mp}}{(n-1)mp}(1-e^{-1/mp})^{-1}.
\label{theta21}
\end{equation}
Combing Eqs. (\ref{barhok}), (\ref{barho}), and (\ref{theta21}),
we find the solution of the density of active nodes when there is
no external stimulus,
\begin{equation}
\rho(1) \sim \frac{1}{n-1}e^{-1/mp}. \label{barho2}
\end{equation}

ii) the uncoupled case, i.e., $p=0$. Combing Eqs. (\ref{barhok})
and (\ref{barho}), we have
\begin{equation}
\label{barho3} \rho(1) = \frac{\lambda}{1 + (n-1)\lambda},
\end{equation}
which is independent of the network's topology, so it is also have
been obtained in other networks, such as one dimensional
case~\cite{Furtado06}. When the external stimulus intensity $r$ is
very small, $\rho(1) \sim r^{-1}$.

\begin{figure}
\includegraphics*[width=\columnwidth]{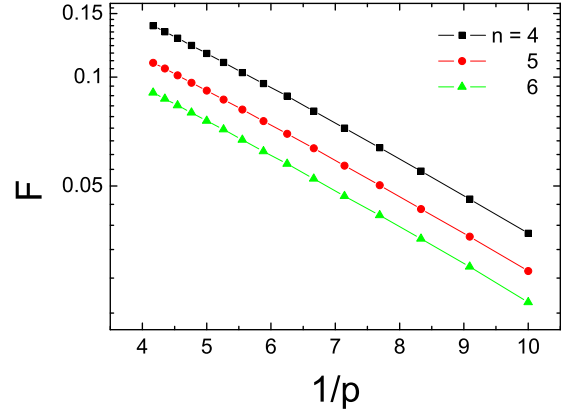}

\caption{In the absence of stimulus, the average activity $F$ as a
function of $1/p$ for BA networks of size $N=10^4$ and
$m_{0}=m=4$. The linear behavior on the semi-logarithmic scale
proves an exponential behavior predicted by Eq.
(\ref{barho2}).}\label{fig1}
\end{figure}

We define the average activity
\begin{equation}
F=\frac{1}{T}\sum_{t=1}^T \rho_t(1), \label{activity}
\end{equation}
where $T$ is a large time window (of the order of $10^4$ time
steps). In the stationary state, it is obviously that $F=\rho(1)$.
To confirm the picture extracted from the above analytic
treatment, we perform numerical simulations on the BA network.
Figure \ref{fig1} shows the behavior of the average activity $F$
in the absence of stimulus. All the plots decays with an exponent
form $F \sim \exp(-c/mp)$, where $c$ is a constant. The numerical
value obtained $c=0.226$ is in good agreement with the theoretical
prediction $c=1/m=0.25$.

\begin{figure}
\includegraphics*[width=\columnwidth]{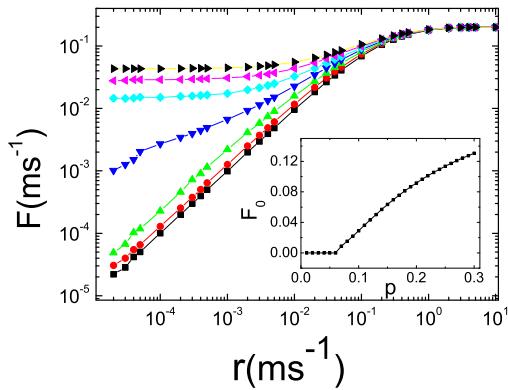}

\caption{Response curves (mean firing rate vs. stimulus rate).
Points represent simulation results with $N=10^4$, $m_{0}=m=4$, $n
= 5$ states and $T = 10^3$~ms, from $p = 0$ to $p=0.12$ (in
intervals of 0.02). These curves are power laws $F \propto
r^\alpha$ with $\alpha = 1$ (subcritical) and $\alpha = 1/2$
(critical). Inset: spontaneous activity $F_0$ vs. branching
probability $p$, the critical point is $p_{c}=0.06$.}\label{fig2}
\end{figure}

In Fig. \ref{fig2}, we show the average activity $F$ versus the
external stimulus rate $r$ for different branching probability
$p$. The inset shows the spontaneous activity $F_0$ versus the
branching probability $p$ in the absence of stimulus ($r = 0$),
there is a critical point $p_c=0.06$, only at $p> p_c$ there is a
self-sustained activity, $F>0$. The curves $F(r)$ could be fitted
by a Hill function, but are not exactly Hill, and there exists the
relation $F\propto r^\alpha$ for the low-stimulus response. It is
found that the Stevens-Hill exponent $\alpha$ changes from $\alpha
= 1$ in the subcritical regime to $\alpha = 0.5$ at criticality.
This important point was also reported by ref. \cite{Kinouchi}
where the network is an Erd\"os-R\'enyi random graph. We also
notice that apparent exponents between 0.5 and 1.0 are observed
\cite{Furtado06} if finite size effects are present, that is, if
$N$ is small.

\begin{figure}
\includegraphics*[width=\columnwidth]{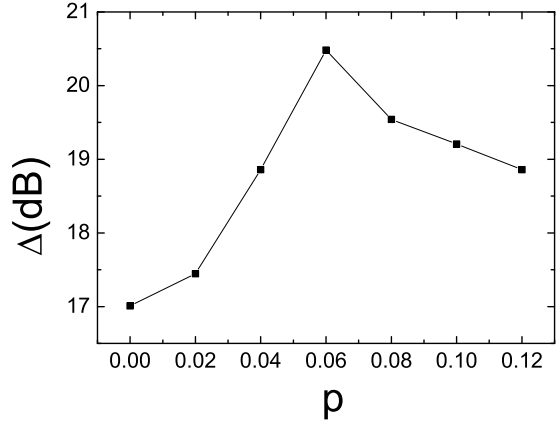}

\caption{Dynamic range vs.  branching probability $p$. The curve
is obtained by calculating the data from Fig.\ref{fig2}. In the
subcritical regime, the dynamic range $\Delta(p)$ increases
monotonically with $p$, In the supercritical regime, $\Delta(p)$
decreases when $p$ increases. There is a maximal range precisely
at the critical point $p_c = 0.06$.}\label{fig3}
\end{figure}

As a function of the stimulus intensity $r$, networks have a
minimum response $F_0$ (= 0 for the subcritical and critical
cases) and a maximum response $F_{max}$ (due to the absolute
nature of the refractory period, $F_{max}=1/n$, which can be
obtained by setting $\lambda=1$ in Eqs. (\ref{barho3})). Ref.
\cite{Kinouchi} defines the dynamic range
$\Delta=10\log(r_{0.9}/r_{0.1})$ as the stimulus interval
(measured in dB) where variations in $r$ can be robustly coded by
variations in $F$, discarding stimuli which are too weak to be
distinguished from $F_0$ or too close to saturation. The range
$[r_{0.1},r_{0.9}]$ is found from its corresponding response
interval $[F_{0.1},F_{0.9}]$, where $F_x = F_0 + x(F_{max}-F_0)$.

Figure \ref{fig3} depicts the dynamic range $\Delta$ versus the
branching probability $p$. In the subcritical regime, the dynamic
range $\Delta(p)$ increases monotonically with $p$, In the
supercritical regime, $\Delta(p)$ decreases when $p$ increases.
There is a maximal range precisely at the critical point. This
result was also found in Erd\"os-R\'enyi random graph
\cite{Kinouchi}. Kinouchi and Copelli explained the result as
\lq\lq In the subcritical regime, sensitivity is enlarged because
weak stimuli are amplified due to activity propagation among
neighbours. As a result, the dynamic range $\Delta(p)$ increases
monotonically with $p$. In the supercritical regime, the
spontaneous activity $F_0$ masks the presence of weak stimuli,
therefore $\Delta(p)$ decreases. The optimal regime occurs
precisely at the critical point\rq\rq \cite{Kinouchi}. We also
perform simulations on other complex networks, such as WS networks
and random SF networks with different $\gamma$, and obtain the
same behavior of the average activity $F$ as it is shown in Fig.
\ref{fig2}, the optimal regime occuring precisely at the critical
point. We state that this phenomenon is the universal behavior of
excitable GHCA model on complex networks and the explanation
provided by Kinouchi and Copelli is also suitable for other types
of complex networks.

\begin{figure}
\includegraphics*[width=\columnwidth]{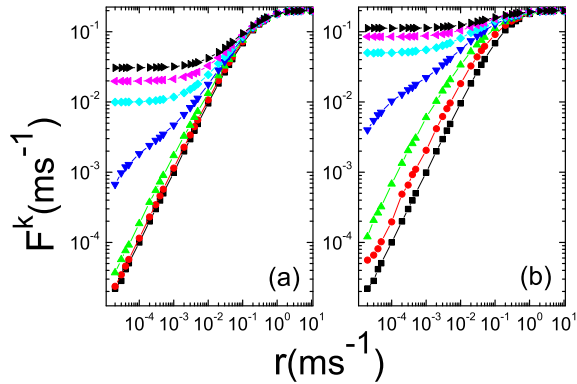}

\caption{Average activity $F^k$ for nodes with given connectivity
$k$ vs. external stimulus rate $r$. (a) for connectivity $k = 4$,
and (b) for connectivity $k = 36$. Simulations are performed with
$N=10^4$, $m_{0}=m=4$, $n = 5$ states and $T = 10^4$~ms, from $p =
0$ to $p=0.12$ (in intervals of 0.02). These curves are power laws
$F^{k} \propto r^\alpha$ with $\alpha = 1$ (subcritical) and
$\alpha = 1/2$ (critical). }\label{fig4}
\end{figure}

In SF networks, there exist strong fluctuations in the
connectivity distribution. It is worthy to investigate the
behavior of the average activity $F^k$ for nodes with given
connectivity $k$. In Fig. \ref{fig4} we plot the quantity $F^k$
versus the external stimulus intensity $r$ for different branching
probability $p$. Figs. \ref{fig4}(a) and (b) correspond to the
node's connectivity $ k = 4$ and $36$, respectively. It is found
that the curves $F^{k}(r)$ have the similar property of $F(r)$
which was shown in Fig. \ref{fig2}, i.e., the Stevens-Hill
exponent $\alpha$ changes from $\alpha = 1$ in the subcritical
regime to $\alpha = 0.5$ at criticality.

\begin{figure}
\includegraphics*[width=\columnwidth]{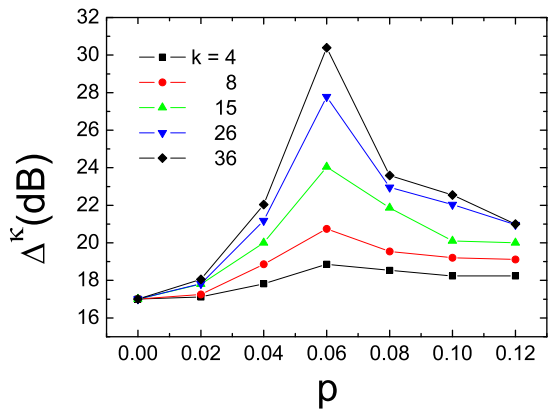}

\caption{ Dynamic range $\Delta^{k}(p)$ of nodes with given
connectivity $k$ vs. branching probability $p$. Points represent
the results calculated from simulation with $N = 10^4$ sites,
$m_{0}=m=4$, $n = 5$ states and $T = 10^4$~ms.  Dynamic range
$\Delta^{k}(p)$ is optimized at the critical point $p_c=0.06$,
although the connectivity $k$ is different .}\label{fig5}
\end{figure}

Finally, we calculate the dynamic range $\Delta^{k}(p)$ of nodes
with given connectivity $k$, and the result is shown in Fig.
\ref{fig5}. The phenomenon that the optimal regime occurs
precisely at the critical point recurs. It is notable that the
optimal dynamic range for nodes with given connectivity $k$
increases with $k$, i.e., for the node with larger connectivity,
its corresponding larger optimal range. One can investigate every
node's optimal dynamic range and calculate their fluctuations,
since the fluctuations of node's optimal dynamic range reflect the
fluctuations of the node's connectivity in the network. To some
extent, we can discovery the topology of the network by
investigating the dynamics on it.

In summary, we have investigated the behavior of the excitable
GHCA model \cite{Greenberg78} on BA networks. We found out that
the curves of the average activity $F$ as a function of the
external stimulus rate $r$ can be fitted by a Hill function, but
are not exactly Hill, and there exists a relation $F\propto
r^\alpha$ for the low-stimulus response. The Stevens-Hill exponent
$\alpha$ changes from $\alpha = 1$ in the subcritical regime to
$\alpha = 0.5$ at criticality. There is a maximal range precisely
at the critical point. We also observed these results numerically
in other kind of complex networks. We conclude that these
phenomena are the universal behaviors of the excitable GHCA model
on complex networks. Due to strong fluctuations in the
connectivity distribution on the BA graph, we calculated the
average activity $F^{k}(r)$ and the dynamic range $\Delta^{k}(p)$
for nodes with given connectivity $k$. The two quantities
$F^{k}(r)$ and $\Delta^{k}(p)$ have the similar behavior as that
of $F(r)$ and $\Delta(p)$, respectively. It is interesting that
nodes with larger connectivity have larger optimal range. This
property could be applied in biological experiment revealing the
network topology.

This work was supported by the Fundamental Research Fund for
Physics and Mathematics of Lanzhou University under Grant No.
Lzu05008. X.-J. Xu acknowledges financial support from FCT
(Portugal), Grant No. SFRH/BPD/30425/2006.

\end{document}